\documentclass[review]{elsarticle}

\journal{Energy}

\usepackage{bm}
\usepackage{amssymb}
\usepackage{mathtools}
\usepackage{siunitx}
\usepackage{fancyref}
\usepackage{pgfplots}
\usepackage{todonotes}
\usepackage{eurosym}
\usepackage{multirow}
\usepackage{xcolor}
\usepackage{tikz} 
\usepackage{hyperref}

\usetikzlibrary{math} 

\DeclareSIUnit{\ct}{ct}
\DeclareSIUnit\year{year}
\DeclareSIUnit{\sieuro}{\mbox{\euro}}

\begin{document}

\begin{frontmatter}
\title{Economic Topology Optimization of District Heating Networks using a Pipe Penalization Approach}

\author[1,2,3]{Yannick Wack\corref{mycorrespondingauthor}}
\cortext[mycorrespondingauthor]{Corresponding author}
\ead{yannick.wack@kuleuven.be}

\author[1,3]{Martine Baelmans}

\author[2,3]{Robbe Salenbien}

\author[1,3]{Maarten Blommaert}

\address[1]{Department of Mechanical Engineering, KU Leuven, Celestijnenlaan 300 box 2421, 3001 Leuven, Belgium}

\address[2]{Flemish Institute for Technological Research (VITO), Boeretang 200, 2400 Mol, Belgium}

\address[3]{EnergyVille, Thor Park, Poort Genk 8310, 3600 Genk, Belgium}

\begin{abstract}
	In the presented study, a pipe penalization approach for the economic topology optimization of District Heating Networks is proposed, drawing inspiration from density-based topology optimization. For District Heating Networks, the upfront investment is a crucial factor for the rollout of this technology. Today, the pipe routing is usually designed relying on a linearization of the underlying heat transport problem. This study proposes to solve the optimal pipe routing problem as a non-linear topology optimization problem, drawing inspiration from density-based topology optimization. The optimization problem is formulated around a non-linear heat transport model and minimizes a detailed net present value representation of the heating network cost. By relaxing the combinatorial problem of pipe placement, this approach remains scalable for large-scale applications. A discrete network topology and near-discrete pipe design is achieved by using an intermediate pipe penalization strategy. For a realistic test case, the proposed algorithm achieves a discrete network topology and near-discrete pipe design that outperforms simple post-processing steps.
\end{abstract}	


\end{frontmatter}



\newcommand*{\folderPath}{}
\newcommand*{\tikzPath}{}
\newcommand*{\imagePath}{}
\newcommand*{\plotPath}{}
\newcommand*{\bibPath}{}
\newcommand*{\mathPath}{}

\ifdefined\unit\else
	\ifdefined\NewCommandCopy
		\NewCommandCopy\unit\si
	\else
	\NewDocumentCommand\unit{O{}m}{\si[#1]{#2}}
	\fi
\fi

\newcommand{\tp}[1]{#1^{\intercal}} 			
\DeclarePairedDelimiter\abs{\lvert}{\rvert} 	
\newcommand{\card}[1]{\lvert#1\rvert}
\newcommand{\infNorm}[1]{\|#1\|_{\infty}}
\newcommand{\Real}[1]{\mathbb{R}^{#1}}
\newcommand{\ve}[1]{\bm{#1}} 	

\newcommand{\gEdge}[3]{#1_{#2#3}}   
\newcommand{\gNode}[2]{#1_{#2}}     

\newcommand{\dirGraph}{G}
\newcommand{\setNodes}{N}
\newcommand{\setEdges}{E}

\newcommand{\pro}{\mathrm{prod}}
\newcommand{\con}{\mathrm{con}}
\newcommand{\rad}{\mathrm{hs}}
\newcommand{\byp}{\mathrm{bp}}
\newcommand{\jun}{\mathrm{jun}}
\newcommand{\pipe}{\mathrm{pipe}}
\newcommand{\operation}{\mathrm{op}}
\newcommand{\Npro}{\setNodes_\pro}
\newcommand{\Ncon}{\setNodes_\con}
\newcommand{\NconF}{\setNodes_{\con,\mathrm{f}}}
\newcommand{\NconR}{\setNodes_{\con,\mathrm{r}}}
\newcommand{\Njun}{\setNodes_\jun}
\newcommand{\NproF}{\setNodes_{\pro,\mathrm{f}}}
\newcommand{\NproR}{\setNodes_{\pro,\mathrm{r}}}

\newcommand{\Epro}{\setEdges_\pro}
\newcommand{\EproF}{\setEdges_{\pro,\mathrm{f}}}
\newcommand{\EproR}{\setEdges_{\pro,\mathrm{r}}}

\newcommand{\Econ}{\setEdges_\con}
\newcommand{\Erad}{\setEdges_\rad}
\newcommand{\Ebyp}{\setEdges_\byp}
\newcommand{\Epipe}{\setEdges_\pipe}
\newcommand{\EpipeF}{\setEdges_{\pipe,\mathrm{f}}}
\newcommand{\EpipeR}{\setEdges_{\pipe,\mathrm{r}}}
\newcommand{\Eop}{\setEdges_\operation}

\newcommand{\n}{n}
\newcommand{\npipes}{\n_{\pipe}}
\newcommand{\defnpipes}{\npipes = \card{\Epipe}}

\newcommand{\gi}{i}
\newcommand{\gj}{j}

\newcommand{\giNode}[1]{\gNode{#1}{\gi}}
\newcommand{\gjNode}[1]{\gNode{#1}{\gj}}
\newcommand{\gijEdge}[1]{\gEdge{#1}{\gi}{\gj}}

\newcommand{\cost}{J}
\newcommand{\designVarskal}{\varphi}
\newcommand{\designVar}{\ve{\designVarskal}}
\newcommand{\stateVar}{\ve{x}}
\newcommand{\topVar}{\ve{\diameter}}
\newcommand{\pentopVar}{\ve{\penDiameter}}

\newcommand{\equalCon}{\ve{c}}
\newcommand{\inEqualCon}{\ve{h}}

\newcommand{\prodInput}{\gamma} 	
\newcommand{\radValve}{\alpha}    
\newcommand{\equalNew}{\tilde{\equalCon}}
\newcommand{\inEqualNew}{\tilde{\inEqualCon}}
\newcommand{\designVarNew}{\tilde{\designVar}}

\newcommand{\radFlowNew}{\tilde{\radValve}}
\newcommand{\prodInputNew}{\tilde{\prodInput}}
\newcommand{\equalConModel}{\equalCon_\mathrm{m}}
\newcommand{\inEqualModel}{\equalCon_\mathrm{s}}
\newcommand{\stateVarModel}{\stateVar_\mathrm{m}}
\newcommand{\stateVarIneq}{\stateVar_\mathrm{s}}

\newcommand{\ALagrangian}{\mathcal{L}}
\newcommand{\LagMultis}{\lambda}
\newcommand{\LagPen}{\mu}
\newcommand{\slack}{s}
\newcommand{\equalConState}{g}

\newcommand{\costFull}{\mathcal{J}}
\newcommand{\costi}[1]{\cost_{\mathrm{#1}}}

\newcommand{\subCAPEX}{CAP}
\newcommand{\subOPEX}{OP}
\newcommand{\subPipe}{pipe}
\newcommand{\subHeat}{h}
\newcommand{\subPump}{p}
\newcommand{\subRev}{rev}

\newcommand{\npv}{\mathrm{NPV}}
\newcommand{\npvConst}{f}
\newcommand{\npvN}{A}
\newcommand{\npvt}{t}
\newcommand{\npvDiscount}{e}
\newcommand{\npvConstCAPEX}{\npvConst_{\mathrm{\subCAPEX}}}
\newcommand{\npvConstOPEX}{\npvConst_{\mathrm{\subOPEX}}}

\newcommand{\npvCost}{C}

\newcommand{\Jpipepol}{\kappa}
\newcommand{\Jpipesmooth}{k}
\newcommand{\cPipe}{\npvCost_{\mathrm{\subPipe}}}

\newcommand{\setDiscreteDiameters}{\hat{D}}
\newcommand{\nDiscretePipes}{m}
\newcommand{\costDiscreteDiameters}{\hat{\npvCost}_{\mathrm{\subPipe}}}
\newcommand{\cHeatCAPEX}{\npvCost_{\mathrm{hC}}}
\newcommand{\cHeatCAPEXi}[1]{\npvCost_{\mathrm{hC},#1}}

\newcommand{\cHeatOPEX}{\npvCost_{\mathrm{hO}}}
\newcommand{\cHeatOPEXi}[1]{\npvCost_{\mathrm{hO},#1}}

\newcommand{\pumpEff}{\eta}
\newcommand{\cPumpOPEX}{\npvCost_{\mathrm{pO}}}
\newcommand{\cPumpOPEXi}[1]{\npvCost_{\mathrm{pO},#1}}
\newcommand{\cPumpCAPEX}{\npvCost_{\mathrm{pC}}}
\newcommand{\cPumpCAPEXi}[1]{\npvCost_{\mathrm{pC},#1}}

\newcommand{\cRev}{\npvCost_{\mathrm{r}}}
\newcommand{\cRevi}[1]{\npvCost_{\mathrm{r},#1}}
\newcommand{\flow}{q}
\newcommand{\pressure}{p}
\newcommand{\temperature}{T}

\newcommand{\Reynolds}{Re}      
\newcommand{\density}{\rho}     
\newcommand{\viscosity}{\mu}    
\newcommand{\spHeatCap}{c_{\mathrm{p}}}

\newcommand{\TOutside}{\temperature_\infty}
\newcommand{\dTinf}{\theta}                 
\newcommand{\heat}{Q}

\newcommand{\length}{L} 
\newcommand{\diameter}{d}
\newcommand{\diameterMin}{\diameter_{\mathrm{min}}}
\newcommand{\diameterDiscrete}{D}
\newcommand{\volume}{V}

\newcommand{\rough}{\epsilon} 
\newcommand{\ratioInsul}{r} 
\newcommand{\condInsul}{\lambda_{\mathrm{i}}} 
\newcommand{\condGround}{\lambda_{\mathrm{g}}} 

\newcommand{\subScriptOuterD}{o}  
\newcommand{\depthPipe}{h} 

\newcommand{\hydrR}{R}  
\newcommand{\frictionfactor}{f} 	
\newcommand{\thermR}{R} 

\newcommand{\inflowID}{a}    
\newcommand{\outflowID}{b}

\newcommand{\valveRhydr}{\zeta}   

\newcommand{\bypValve}{\alpha}     

\newcommand{\lmtd}{\mathrm{LMTD}}
\newcommand{\heaterCoef}{\Phi}	
\newcommand{\heaterExp}{n} 
\newcommand{\THouse}{\dTinf_{\textrm{house}}}

\newcommand{\Qdemand}{Q_{\mathrm{d}}}
\newcommand{\Qdemandi}[1]{Q_{\mathrm{d},#1}}
\newcommand{\DemandSatisfaction}{S}

\newcommand{\prodInputi}[1]{\prodInput_{#1}} 	

\newcommand{\prodTemp}{\Theta}
\newcommand{\prodTempv}{\ve{\Theta}}
\newcommand{\prodTempi}[1]{\Theta_{#1}} 

\newcommand{\pen}{\xi} 
\newcommand{\penDirection}{a}
\newcommand{\penDiameter}{\bar{\diameter}}
\newcommand{\setDiameters}{\mathcal{S}}
\newcommand{\relu}{f}


\newcommand{\defSetPipes}{\forall \gi\gj \in \Epipe}
\newcommand{\defSetRad}{\forall \gi\gj \in \Erad}
\newcommand{\defSetByp}{\forall \gi\gj \in \Ebyp}
\newcommand{\defSetCon}{\forall \gi\gj \in \Erad \cup \Ebyp}
\newcommand{\defSetProd}{\forall \gi\gj \in \Epro}

\newcommand{\defInflow}{\inflowID=(\gi,n) \in \setEdges}
\newcommand{\defOutflow}{\outflowID=(n,\gj) \in \setEdges}

\newcommand{\topVarBoxConstraints}[1]{\diameterDiscrete_{0}\leq#1\leq\diameterDiscrete_{N}}
\newcommand{\opVarBoxConstraints}[1]{ 0\leq #1 \leq 1}
\newcommand{\defFlow}{\ve{\flow} \in \Real{\card{\setEdges}}}
\newcommand{\defPressure}{\ve{\pressure} \in \Real{\card{\setNodes}}}
\newcommand{\defTemp}{\ve{\dTinf} \in \Real{\card{\setNodes\cup\setEdges}}}

\newcommand{\defStateVar}{\ve{\stateVar} = \tp{\left[\ve{\flow},\ve{\pressure},\ve{\dTinf}\right]} \in \mathbb{R}^{2\card{\setNodes}\card{\setEdges}}}

\newcommand{\defDesignVar}{\designVar=\tp{\left[\ve{\radValve},\ve{\prodInput}\right]} \in \Real{\card{\Eop}}}
\newcommand{\setDefDiscreteDiameters}{\{\diameterDiscrete_{0},\dots,\diameterDiscrete_{\nDiscretePipes}\}}
\newcommand{\defTopVar}{\topVar\in\setDefDiscreteDiameters^{\card{\Epipe}}}

\newcommand{\defModelConstraints}{\equalCon\left(\topVar, \designVar,\stateVar\right)}

\definecolor{colorplot1}{RGB}{0,64,122}		
\definecolor{colorplot2}{RGB}{178,34,34}  	
\definecolor{colorplot3}{RGB}{34,139,34}  	

\section{Introduction}

In recent years, the field of topology optimization has seen widespread application in structural mechanics, fluid flow, or heat transfer problems. Here, challenging large-scale integer programming problems are solved by initially relaxing the integer variables and later steering it towards discrete solutions \cite{Deaton2014}. Due to these relaxations, the approach remains scalable, and large optimal design problems can be solved. In contrast, to optimize network topologies, combinatorial optimization approaches have traditionally been used (e.g. in transmission expansion planning \cite{Lumbreras2016a}). In these applications, networks can often be accurately described as linear optimization problems, which can be solved efficiently using Mixed Integer Programming (MIP) solvers. Coming from this network tradition, the topology of heating networks has mostly been optimized using MIP solvers. Examples include Resimont et al. \cite{Resimont2021}, Mertz et al.\cite{Mertz2016} or Dorfner \& Hamacher \cite{dorfner2014large}.

District Heating Networks (DHNs) are a network technology connecting heat demands and supplies through a network of insulated pipes carrying hot water. Due to its ability to connect a multitude of different renewable heat sources and provide heat to districts and entire cities, it is considered one of the core technologies to enable carbon-neutral space heating \cite{OECD/IEA2019}. In DHNs, the typically high upfront investment cost of groundworks and piping is a core decision variable for the feasibility of a development project. Therefore in the planning phase it is crucial to economically optimize a heating network while achieving an optimal discrete topology and pipe design.

The transformation of modern DHNs towards multi-source, low-temperature networks \cite{Lund2021}, breaks the assumptions of most linear heating network models that are used to aid in the design. To account for different feasible temperature levels of supply and demand sites, and to accurately  model heat losses, a non-linear representation of the network physics is necessary. The routing choice for DHN pipes as well as the limitations of pipe diameters to manufacturer catalogues further poses the optimal topology problem of DHNs a discrete optimization problem. The need to solve a Mixed Integer Non-Linear Program (MINLP) containing many discrete variables renders most classic combinatorial optimization approaches inefficient. Attempts to solve this MINLP using combinatorial solvers have notably been made by Mertz et al. \cite{Mertz2016}. The computational complexity of solving said MINLP leaves potential to explore the application of topology optimization methods to DHNs. A first step in this direction was published by the authors of this paper in Blommaert et al. \cite{Blommaert2020b}. This paper focused on defining a non-linear District Heating Model and developing a fast optimization algorithm by using adjoint gradients on a simplified cost function for heating network design. Additionally, a first proposal to achieve a near-discrete design was made. 

The most popular approach to ensure near-discrete design in density-based topology optimization for PDE constrained optimization is the Solid Isotropic Material with Penalization (SIMP) approach. Here, the discrete variable is replaced with a continuous variable, and is steered towards a discrete solution using implicit penalization \cite{Deaton2014}. By applying density-based topology optimization for the first time to DHNs, we propose in this paper a SIMP-like multi-material penalization approach to economically optimize the topology of DHNs while achieving a near-discrete topology and pipe design. 

Building on the previous paper by the authors \cite{Blommaert2020b}, three new crucial contributions are elaborated here. First, the optimal design problem is reformulated as an economical optimization problem, allowing for accurate design studies of future DHN development projects. This reformulation is laid out in Section \ref{sec:Cost}. Second, as technological (state) constraints play an important role in DHN optimization, the optimization algorithm is rewritten using an Augmented Lagrangian approach (See Section \ref{sec:AugLag}). Third, we propose a topology optimization approach using pipe diameter penalization for optimal DHN design. It is based on a penalization technique inspired by the SIMP method of Bends{\o}e \cite{Bendsoe1989a} and its multi-material application by Zou and Saitou \cite{Zuo2017}.  

\section{The topology optimization problem, a compromise between scalability and physical accuracy}\label{sec:OptProblem}
\begin{figure}[h]
	\centering
	\includegraphics{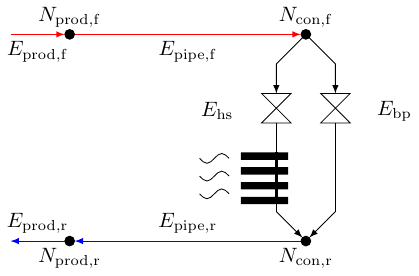}
	\caption{Minimal illustration of the components of a District Heating network and their graph representation.}
	\label{fig:DHNcomponents}
\end{figure}

DHNs are a network technology, connecting heat demands and supply by a network of water carrying insulated pipes. They therefore can be represented in a directed graph $\dirGraph(\setNodes,\setEdges)$, with
$\setNodes$ the set of all nodes and $\setEdges$ the set of all edges in the graph. 
We can further subdivide the set of nodes $\setNodes$ into three subsets $\Npro \cup \Ncon \cup \Njun =  \setNodes$, denoting the producer, consumer, and pipe junctions. Similarly, the set of edges $\setEdges$ is partitioned into the subsets  $\Epro \cup \Erad \cup \Ebyp \cup \Epipe =  \setEdges $, denoting the producer, consumer heating system, consumer bypass edges, and edges representing (potential) pipes, respectively. To differentiate between feed and return network, all node subsets $\setNodes$ as well as the edge subsets $\Epipe$ and $\Epro$ can be further subdivided into feed and return components (e.g. $\Epro = \EproF \cup \EproR$). For simplicity of further notation, edges representing the consumer heating system and bypasses are grouped as consumer edges $\Econ = \Erad \cup \Ebyp$, and together with the producer feed edges form the set of operational edges $ \Eop =\EproF \cup \Econ$. The set definition of different DHN components is illustrated in figure \ref{fig:DHNcomponents}.  We can now use the cardinality to determine the number of components in each subset, e.g. the number of pipes in the network: $n_{\mathrm{\pipe}}=\card{\Epipe}$.
In the following sections, we now denote a network node as $n \in \setNodes$ and a directed edge going from node $\gi$ to node $\gj$ as $(\gi,\gj) \in \setEdges$, or succinctly as $\gi\gj \in \setEdges$ or even more compactly as $a \in \setEdges$ \cite{Blommaert2020b}. 

Defining the optimal topology and design problem of a DHN as a mathematical optimization problem requires the definition of design variables that are to be chosen in an optimal way. In the case of DHNs these design variables contain the pipe diameters $\topVar\in\setDiscreteDiameters^{\npipes}$, with $\defnpipes$ being the number of possible pipes in the network. The pipe diameters in $\topVar$ are chosen from a set of available discrete pipe diameters $\setDiscreteDiameters=\setDefDiscreteDiameters$,  Here $\diameterDiscrete_{0}$ represents the choice of not placing a pipe. Therefore $\topVar$ acts as the topological variable. Furthermore, an operational design variable $\defDesignVar$ is defined, containing the radiator and bypass valve setting $\ve{\radValve}\in \Real{\card{\Econ}}$ as well as the normalised producer inflow $\ve{\prodInput}\in\Real{\card{\EproF}} $. To represent the physical state of a given network, a vector of physical variables $\defStateVar$ is defined. It contains the flow rates $\defFlow$, nodal pressures $\defPressure$ and nodal and pipe exit  temperature $\defTemp$. The temperature $\ve{\dTinf} = \ve{\temperature} - \TOutside$ is defined as the difference between the absolute water temperature $\ve{\temperature}\in \Real{\card{\setNodes\cup\setEdges}}$ and the outside air temperature $\TOutside$. 

Now the topology optimization problem for DHNs can be posed as a generic optimization problem of the form:

\begin{equation} \label{eq:opt}
	\begin{aligned} \min_{\topVar,\designVar,\stateVar}& \qquad
		\costFull \left(\topVar, \designVar,\stateVar\right) &\textrm{cost function}\\
		s.t.& \qquad \equalCon(\topVar,\designVar,\stateVar) = 0, &\textrm{model equations}\\
		& \qquad \inEqualCon(\topVar,\designVar,\stateVar) \leq 0, &\textrm{state constraints} \\
		& \qquad \topVar\in\setDefDiscreteDiameters^{\npipes} & \\
		& \qquad \opVarBoxConstraints{\designVar}.&
	\end{aligned}
\end{equation}

The three main components of this optimization problem will be elaborated in the following sections. First a cost function $\costFull \left(\topVar, \designVar,\stateVar\right)$ is defined in section \ref{sec:Cost}. Then a set of model equations $\defModelConstraints$, describing the networks physics is defined in section \ref{sec:Model}, and finally additional state constraints $\inEqualCon\left(\topVar, \designVar,\stateVar\right)$ are formulated in section \ref{sec:stateConst}. 

\subsection{Cost function}\label{sec:Cost}
In contrast to the previous publication by Blommaert et al. \cite{Blommaert2020b}, this paper aims at a detailed economic optimization. As the current bottleneck for the development of DHNs is their high upfront investment cost, it is important to accurately describe and optimize the economic cost of a DHN project. To account for upfront and future cash flows, a cost function is defined that maximizes the net present value $NPV$ \cite[p.~273]{CorporateFinance2017} of a planned Heating Network:

\begin{align} \label{eq:totalCost}
	\costFull\left(\topVar,\designVar,\stateVar\right) &= -\npv\left(\topVar,\designVar,\stateVar\right) \nonumber\\
	&= \costi{\subPipe,\subCAPEX}\left(\topVar\right) + \costi{\subHeat,\subCAPEX}\left(\designVar,\stateVar\right)  \nonumber\\
	&+ \npvConstOPEX \left[\costi{\subHeat,\subOPEX}\left(\designVar,\stateVar\right) + \costi{\subPump,\subOPEX}\left(\designVar,\stateVar\right)\right. \nonumber \\
	&-\left.\costi{\subRev}\left(\designVar,\stateVar\right) \right] \,,
\end{align}

with

\begin{equation} \label{eq:npvDiscount}
	\npvConstOPEX = \sum_{\npvt=1}^{\npvN} \frac{1}{\left(1+\npvDiscount\right)^\npvt} \, .
\end{equation}
Here we assume that cash flows over the investment horizon $\npvN = 30~\mathrm{years}$ remain constant. A discounting rate of $\npvDiscount=5\%$ is assumed.

The investment cost of piping $\costi{\subPipe,\subCAPEX}$ is approximated with a linear interpolation of the catalogue cost per meter $\costDiscreteDiameters=\{\npvCost_{\mathrm{\subPipe},1},\dots,\npvCost_{\mathrm{\subPipe},\nDiscretePipes}\}$ for the set of available discrete pipe diameters $\setDiscreteDiameters=\{\diameterDiscrete_1,\dots,\diameterDiscrete_{\nDiscretePipes}\}$, resulting in the interpolation coefficients $\Jpipepol_1$ and $\Jpipepol_0$. The pipe investment cost then reads:
\begin{equation} \label{eq:Jpipe}
	\costi{\subPipe,\subCAPEX}\left(\topVar\right) = \sum_{\gi\gj \in \Epipe}\left(\Jpipepol_1  \gijEdge{\diameter} + \frac{1}{2}\bar{\Jpipepol}_0(\gijEdge{\diameter}) \right)\gijEdge{\length}\,. 
\end{equation}
To smoothly account for the cost reduction of topological changes ($\costi{\subPipe,\subCAPEX}\left(\diameterMin\right) :=0$), the fixed piping cost $\bar{\Jpipepol}_0$ is modelled similar to Pizzolato et al. \cite{Pizzolato2018} using
\begin{equation}
	\bar{\Jpipepol}_0(\gijEdge{\diameter})= \Jpipepol_0 \left(\frac{2}{(1+\exp(-\Jpipesmooth\left(\gijEdge{\diameter}-\diameterMin\right))})-1\right)\,,
\end{equation}
$\defSetPipes$, with a steepness of $\Jpipesmooth = 600$ and a minimum pipe diameter of $\diameterMin = 1\si{\milli \meter}$. 

The investment cost for building heat production plants is calculated using
\begin{equation}
	\costi{\subHeat,\subCAPEX}\left(\stateVar\right) = \sum_{\gi\gj \in \EproF}\cHeatCAPEXi{\gi\gj}\gijEdge{\flow} \,\gijEdge{\dTinf}\, \density \spHeatCap  \, ,
\end{equation}
with $\cHeatCAPEX$ being the capacity price of heat production in $\si{\sieuro\per{\watt}}$. Here we denote the water density with $\density$ and the specific heat capacity of water with $\spHeatCap$. The operational cost in \euro/year is calculated using
\begin{equation}
	\costi{\subHeat,\subOPEX}\left(\stateVar\right) = \frac{8760\si{\hour}}{\si{year}}\sum_{\gi\gj \in \EproF}\cHeatOPEXi{\gi\gj}\gijEdge{\flow} \,\gijEdge{\dTinf}\, \density \spHeatCap \, ,
\end{equation}
with the unit price of heat $\cHeatOPEX$ in $\si{\sieuro\per{\watt\hour}}$. The operational cost of pumps at the heat production sites is computed with
\begin{equation}
	\costi{\subPump,\subOPEX}\left(\stateVar\right) = \frac{1}{\pumpEff} \,\frac{8760\si{\hour}}{\si{year}}\sum_{\gi\gj \in \EproF}\cPumpOPEXi{\gi\gj}\left(\gjNode{\pressure}-\pressure_{a}\right) \gijEdge{\flow} \, ,
\end{equation}
where $\pressure_{a}$ with $a\in \NproR$ represents the corresponding pressure at the return node of a producer. The unit pumping price is defined by the electricity price $\cPumpOPEX$ in $\si{\sieuro\per{\watt\hour}}$ and the pump efficiency is given by $\pumpEff$. The investment cost for these pumps is calculated using 
\begin{equation}
	\costi{\subPump,\subCAPEX}\left(\stateVar\right) = \frac{1}{\pumpEff}\sum_{\gi\gj \in \EproF}\cPumpCAPEXi{\gi\gj}\left(\gjNode{\pressure}-\pressure_{a}\right) \gijEdge{\flow} \, ,
\end{equation}
with the pump capacity cost $\cPumpCAPEX$ in $\si{\sieuro\per{\watt}}$. To account for revenue by selling heat to the connected consumers, the revenue cash flow 
\begin{equation}
	\costi{\subRev}\left(\stateVar\right) = \sum_{\gi\gj \in \Erad}\cRevi{\gi\gj}\gijEdge{Q}\SI{8760}{\hour\per\year} \, ,
\end{equation}
is introduced, with a heat selling price $\cRev$ and the heat transferred to a house $\gijEdge{Q}$ defined by equation \ref{eq:heaterConservation}.

\subsection{A nonlinear DHN model}\label{sec:Model}
In contrast to most district heating optimization studies, the optimization model in this paper attempts to accurately capture the flow and heat transfer physics within the network. Therefore, a set of non-linear model equations  $\equalCon(\topVar,\designVar,\stateVar)=0$ for the thermal and hydraulic transport problem will be defined in the following section.
The majority of the network models used in this study were previously established in Blommaert et al. \cite{Blommaert2020b}. For consistency, they are briefly repeated in this section. Some changes were made to increase the stability of model and optimization convergence, which will be further detailed in this section. 

\subsubsection{Pipe model}
In DHNs water is generally used as the carrier fluid, with a constant density $\density$, dynamic viscosity $\viscosity$, and specific heat capacity $\spHeatCap $. Temperature-dependence of these fluid properties is neglected.
In the example elaborated in section \ref{sec:results}, the values will be taken as corresponding to a water temperature of $60 \si{\degreeCelsius}$.
To model the momentum equations over a pipe, we use the empirical Darcy-Weisbach equation, modelling the viscous pressure drop in incompressible flow as a function of the volumetric flow rate $\gijEdge{\flow}$ through a pipe $(\gi,\gj)$ with length $\gijEdge{\length}$: 
\begin{equation} \label{eq:pipeMomentum}
	(\giNode{\pressure} - \gjNode{\pressure}) =  \gijEdge{\frictionfactor} \frac{8\density\gijEdge{\length}}{\gijEdge{\diameter}^5\pi^2}\abs{\gijEdge{\flow}}\gijEdge{\flow} \,,\quad \defSetPipes,
\end{equation}

\begin{equation}
	\mathrm{with} \quad	\gijEdge{\frictionfactor} = 0.3164\left(\Reynolds\right)^{-\frac{1}{4}} \,,\quad \defSetPipes.
\end{equation}
In contrast to Blommaert et al. \cite{Blommaert2020b}, the Darcy friction factor $\gijEdge{\frictionfactor}$ is modelled with the Blasius correlation \cite{BlasiusH.1913}. Here $\gijEdge{\diameter}$ denotes the inner diameter of the pipe and $\Reynolds$ the Reynolds number, defined as
$\Reynolds = \frac{4 \density \abs{\flow}}{(\pi \viscosity \diameter)}$. The non-differentiability of $\abs{\flow}$ at $\flow=0$ is regularized using a cubic fit.

Next, the heat loss of an insulated pipe, installed underground is modelled similar to Van der Heijde et al. \cite{van2017dynamic}. Let us consider $\giNode{\dTinf}$ to be the temperature difference at the node $\gi$, at which the flow enters the pipe $\gi\gj$, and $\gijEdge{\dTinf}$ at the pipe exit. The pipe exit temperature $\gijEdge{\dTinf}$, due to heat loss to the environment is given by
\begin{equation}\label{eq:pipeEnergy}
	\gijEdge{\dTinf} = \giNode{\dTinf} \exp{\left(\frac{-\gijEdge{\length}}{\density \spHeatCap \abs{\gijEdge{\flow}}{\gijEdge{\thermR}}}\right)}, \quad \defSetPipes,
\end{equation}

with $\gijEdge{\thermR}$ the thermal resistance per unit pipe length between the water and the environment.
For a pipe with outer insulation casing diameter $\diameter_{\subScriptOuterD,\gi\gj}$ that is assumed to be bigger than the inner diameter $\diameter_{\gi\gj}$ by a fixed ratio, i.e. $\ratioInsul=\frac{\diameter_{\subScriptOuterD,\gi\gj}}{\gijEdge{\diameter}}  $, the combined thermal resistance of pipe and soil per unit length is \cite{Wallenten1991}
\begin{equation}
	\gijEdge{\thermR} = \frac{\ln(4 \depthPipe/(\ratioInsul \gijEdge{\diameter}))}{2 \pi \condGround} +\frac{\ln{\ratioInsul}}{2 \pi \condInsul},
\end{equation}
with $\condInsul$ and $\condGround$ the thermal conductivity of the insulation and the surrounding ground, respectively, and $\depthPipe$ the depth at which the pipe is buried.

\subsubsection{Pipe junction model}
All nodes in the network represent pipe junctions. For the incompressible flow under consideration conservation of mass reduces to conservation of the flow rate $\flow$ : 
\begin{equation} \label{eq:NodeContinutiy}
	\sum_{\inflowID} \flow_{\inflowID} - \sum_{\outflowID} \flow_{\outflowID} = 0,
\end{equation}
where $\flow_\inflowID$ with $\defInflow$ denotes the flow of incoming edges and $\flow_\outflowID$ with $\defOutflow$ the flow of outgoing edges of a Node $n \in \setNodes$.

Similarly, the temperatures in the node can be determined by conservation of the convected energy. Within the junction, perfect mixing of the incoming flows is assumed. So outgoing flows leave at the node temperature $\dTinf_n$. Note that no assumption is made on the direction of the flows and that depending on the sign of the flow rate $q$ in the directed edges connected to the node, the flow will either enter or leave the junction. Energy conservation can thus be formulated as
\begin{equation}\label{eq:NodeEnergy}
	\begin{aligned}
		& \sum_{\inflowID} \left(\max(\flow_\inflowID,0) \, \dTinf_\inflowID +\min(\flow_\inflowID,0) \,\dTinf_n\right)- \\& \sum_{\outflowID} \left(\max(\flow_\outflowID,0)\,\dTinf_n +\min(\flow_{\outflowID},0)\, \dTinf_{\outflowID}\right) = 0, \quad \forall n \in \setNodes\,
	\end{aligned}
\end{equation}
where again $\flow_\inflowID$ with $\defInflow$ denotes the flow of incoming edges and $\flow_\outflowID$ with $\defOutflow$ the flow of outgoing edges of a Node $n \in \setNodes$.

\subsubsection{Consumer model}
Here a basic model to estimate the heat transferred to the consumer is introduced. Following from the steady state assumption we model the consumer substation and heating system jointly as depicted in figure \ref{fig:DHNcomponents}. 

Both bypass and heating system have a control valve $\gijEdge{\radValve} \in \left[0,1\right],\defSetCon$ to regulate the flow. The pressure drop over both edges is assumed to be in the form
\begin{equation} \label{eq:radiatorMomentum}
	\giNode{\pressure} - \gjNode{\pressure} = \gijEdge{\valveRhydr} \frac{\gijEdge{\flow}}{\gijEdge{\radValve}}, \quad \defSetCon
\end{equation}
with $\gijEdge{\valveRhydr}$ a constant determined from nominal network operating conditions \cite{pirouti2013}. 

Conservation of energy in the heating system leads to
\begin{equation}\label{eq:heaterConservation}
	\density \spHeatCap \gijEdge{\flow}(\giNode{\dTinf} - \gijEdge{\dTinf}) = \gijEdge{\heat}, \quad \defSetRad,
\end{equation}
with $\gijEdge{\heat}$ the heat transferred to the house through the heating system.
The latter is modelled with the characteristic equation for radiators \cite{Ashrae2014c,Heizungstechnik2014}
\begin{equation}\label{eq:characteristicRadiator}
	\gijEdge{\heat} = \gijEdge{\heaterCoef} \left(\lmtd\left(\giNode{\dTinf} - \THouse,\gijEdge{\dTinf} - \THouse\right)\right)^{\gijEdge{\heaterExp}},
\end{equation}
in contrast to Blommaert et al. \cite{Blommaert2020b}, using the $\lmtd$ approximation by Chen \cite{Chen1987} to improve conditioning:
\begin{equation}
	\lmtd\left(\Delta\dTinf_{\mathrm{A}},\Delta\dTinf_{\mathrm{B}}\right)\approx\left(\Delta\dTinf_{\mathrm{A}}\Delta\dTinf_{\mathrm{B}}\left(\frac{\Delta\dTinf_{\mathrm{A}}+\Delta\dTinf_{\mathrm{B}}}{2}\right)\right)^{\frac{1}{3}}\,.
\end{equation}
Here, $\THouse$ is the temperature difference between the indoor and the environment at the house. 
Values of the coefficients $\gijEdge{\heaterCoef}$ and $\gijEdge{\heaterExp}$ are tabulated for individual radiators, according to the EN 442-2 norm \cite{DIN442}. The bypass edges on the other hand are assumed to be free of heat losses, i.e. \begin{equation}\gijEdge{\dTinf} = \giNode{\dTinf} \quad \defSetByp\end{equation}
\subsubsection{Producer model}
In the producer edges, a fixed input flow $\prodInput$ is imposed as boundary condition for this system of equations. In addition, a given temperature $\prodTemp$ is imposed for the heat source. This leads to 
\begin{equation} \label{eq:12}
	\gijEdge{\flow} = \prodInputi{\gi\gj},\quad \gijEdge{\dTinf} = \prodTempi{ij} \quad \forall \gi\gj \in 
	\EproF.
\end{equation}
To uniquely define the pressures throughout the network, a reference pressure is imposed in one of the producer return nodes.

\subsection{Additional state constraints}\label{sec:stateConst}
In addition to satisfying the physical model defined in section \ref{sec:Model}, technological constraints $\inEqualCon(\topVar,\designVar,\stateVar) \leq 0$ have to be defined to ensure that a useful optimization problem is solved. For this study, it is required that the heat demand $\Qdemandi{\gi\gj}\,\defSetRad$ is satisfied for all consumer within a margin of $\pm 5 \%$. This constraints can be formulated as:
\begin{align}\label{eq:heatSatisfaction}
	\pm \left(\frac{\gijEdge{Q}-\Qdemandi{\gi\gj}}{\Qdemandi{\gi\gj}}\right) - 0.05  \leq 0 \,, \quad \defSetRad.
\end{align}

\section{Methodology}
Now that the topology optimization problem for Heat Networks has been defined, a methodology is proposed to solve this non-linear discrete optimization problem. Assuming a superstructure of possible pipe connections constituted by the street network of a neighbourhood in question, a choice has to made whether a pipe is placed, and if so, which available diameter is chosen. The set of available pipe diameters is defined as $\setDiscreteDiameters=\setDefDiscreteDiameters$, from which the pipe diameter $\topVar$ is chosen. To avoid resorting to combinatorial optimization techniques, similar to density-based topology optimization methods, this problem is reformulated as a continuous optimization problem:
\begin{equation} \label{eq:optContinuous}
	\begin{aligned} \min_{\topVar,\designVar,\stateVar}& \qquad
		\costFull \left(\topVar, \designVar,\stateVar\right) \\
		s.t.& \qquad \equalCon(\topVar,\designVar,\stateVar) = 0, \\
		& \qquad \inEqualCon(\topVar,\designVar,\stateVar) \leq 0,  \\
		& \qquad \topVarBoxConstraints{\topVar},   \\
		& \qquad \opVarBoxConstraints{\designVar}.
	\end{aligned}
\end{equation}
The algorithmic steps taken to solve this continuous optimization problem and achieving near-discrete designs are described in the following sections.

\subsection{Achieving near-discrete design: a SIMP-like penalization approach}\label{sec:penalization}
\begin{figure}
	\includegraphics{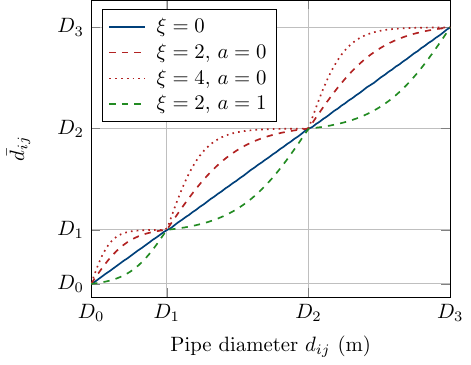}
	\caption{Intermediate diameter penalization for multiple available discrete pipe diameters $\diameter_{\gi\gj} \in \{\diameterDiscrete_{0},\dots,\diameterDiscrete_{3}\} $, for different penalization parameters $\pen$ and directions $\penDirection$.}
	\label{fig:pen}
\end{figure}

By reformulating the topology optimization problem in the aforementioned continuous way, the need arises to steer the design towards a discrete solution. This is typically done by using penalization techniques \cite{Deaton2014}. With the need to pick the optimal diameter from a set of available discrete diameters, this problem strongly resembles multi-material topology optimization problems. These types of problems are often solved by introducing multiple density variables \cite[p.120]{Bendsoe2003}. To reduce the amount of topology variables, we propose to penalize intermediate diameters between available pipes with a multi-material SIMP like approach, similar to the ordered SIMP interpolation by Zou and Saitou \cite{Zuo2017}.  In this approach, normalized densities of multiple materials are sorted by their elastic modulus, to then be described by a single density variable. The sum of normalized densities, or here diameters, can be written as:  
\newcommand{\deltaD}{\Delta \diameterDiscrete}
\begin{equation}\label{eq:penalization}
	\penDiameter_{\gi\gj}\left(\diameter_{\gi\gj},\pen,\penDirection\right)=\sum_{k=0}^{N}\deltaD_{k} \min\left(\max\left(\Pi\left(\diameter_{\gi\gj}\right)
	,0\right),1\right) \,,
\end{equation}
with
\begin{equation}
	\Pi\left(\diameter_{\gi\gj},\pen,\penDirection\right)=\begin{cases}
		\frac{\tanh\left(\pen \frac{\diameter_{\gi\gj}-\diameterDiscrete_{k}}{\deltaD_{k}}-\penDirection\right)}{\tanh\left(\pen\right)}+\penDirection & \text{if $\pen > 0$} \\
		\frac{\diameter_{\gi\gj}-\diameterDiscrete_{k}}{\deltaD_{k}} & \text{if $\pen = 0$}
	\end{cases}\,,
\end{equation}
where $\defSetPipes $ and $\deltaD_{k} = \diameterDiscrete_{k+1} - \diameterDiscrete_{k}$. An illustration of this interpolation can be found in figure \ref{fig:pen}. Intermediate diameters between the discrete options are penalized using a $\tanh$ function, though it is equally possible to use a power law, as is common in the SIMP approach. For this penalization, the parameter $ \pen \in \mathbb R_{\geq}$ controls the steepness, while the direction of penalization is controlled with $\penDirection \in \{0,1\}$.

\begin{figure}[h]
	\includegraphics{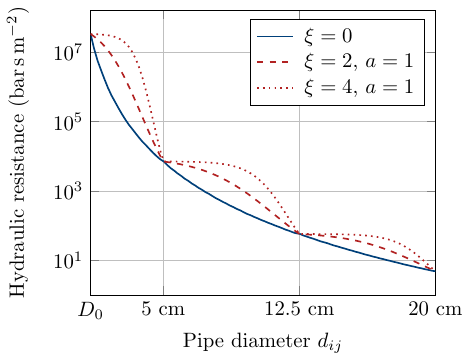}
	\caption{Penalization of intermediate diameters through the hydraulic resistance of pipes for different penalization parameters $\pen$ and a set of available pipe diameters $\diameter_{\gi\gj} \in \{\diameterDiscrete_{0},\SI{5}{\centi\meter},\SI{12.5}{\centi \meter}, \SI{20}{\centi \meter}\}$.}
	\label{fig:penHydr}
\end{figure}

To achieve a penalization of intermediate diameters, similar to the material property interpolation in the SIMP approach, the pipe diameter variable $\ve{\diameter}$ in the  optimization problem is substituted with the penalized diameter $\ve{\penDiameter}\left(\ve{\diameter},\pen,\penDirection\right)$. In the model equations, this leads to an increased hydraulic friction  $\gijEdge{\frictionfactor}$ for $\penDirection=1$ (compare equation \ref{eq:pipeMomentum}),
\begin{equation}
	\gijEdge{\frictionfactor} = 0.3164\left(\frac{4 \density \abs{\flow}}{(\pi \viscosity \penDiameter)}\right)^{-\frac{1}{4}} \,\quad \defSetPipes, \nonumber
\end{equation}
subsequently increasing the hydraulic resistance for intermediate pipe diameters as illustrated in figure \ref{fig:penHydr}.

\begin{figure}[h]
	\includegraphics{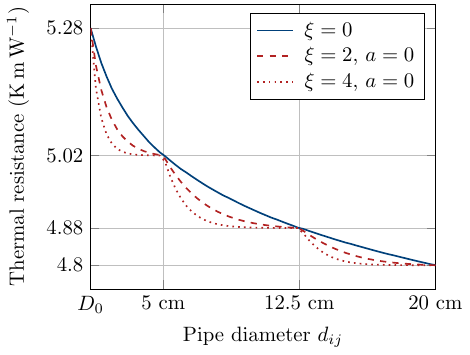}
	\caption{Penalization of intermediate diameters through the thermal resistance of the pipe insulation for different penalization parameters $\pen$ and a set of available pipe diameters $\diameter_{\gi\gj} \in \{\diameterDiscrete_{0},\SI{5}{\centi\meter},\SI{12.5}{\centi \meter}, \SI{20}{\centi \meter}\}$.}
	\label{fig:penTherm}
\end{figure}

Penalizing the pipe diameter in the pipe energy equations for $\penDirection=0$, leads to a decreased thermal resistance $\gijEdge{\thermR}$ of the pipe insulation (compare equation \ref{eq:pipeEnergy} ) for intermediate diameters:
\begin{equation}
	\gijEdge{\thermR} = \frac{\ln(4 \depthPipe/(\ratioInsul \gijEdge{\penDiameter}))}{2 \pi \condGround} +\frac{\ln{\ratioInsul}}{2 \pi \condInsul} \quad \defSetPipes. \nonumber
\end{equation}
This decreasing thermal resistance is illustrated in figure \ref{fig:penTherm}.

\begin{figure}[h]
	\includegraphics{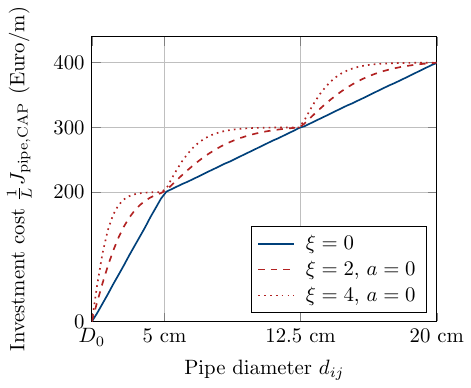}
	\caption{Penalization of intermediate diameters in the investment cost $\frac{1}{\length}\costi{\subPipe,\subCAPEX}$ for different penalization parameters $\pen$ and a set of available pipe diameters $\diameter_{\gi\gj} \in \{\diameterDiscrete_{0},\SI{5}{\centi\meter},\SI{12.5}{\centi \meter}, \SI{20}{\centi \meter}\}$.}
	\label{fig:penInv}
\end{figure}

Similar, a direct penalization is achieved in the pipe investment cost (compare equation \ref{eq:Jpipe} for $\penDirection=0$) and is illustrated in figure \ref{fig:penInv}. These penalizations render intermediate diameters less interesting for the optimizer through increased total costs. In this substitution, the direction of penalization $\penDirection$ is chosen in such a way as to make intermediate diameters unfavourable in the optimization. 

Introducing a high penalization $\pen$, causes an ill-conditioning of the optimization problem, that can hinder convergence. To avoid this ill-conditioning, a reformulation of the initial optimization problem (equation \ref{eq:optContinuous}) as a partially-reduced space formulation is proposed. The detailed reformulation can be found in appendix \ref{secA1}. It leads to a new system of model equations $\tilde{\equalCon}\left(\pentopVar,\designVarNew,\stateVar\right) = 0$ with the new design variable vector $\designVarNew = \tp{\left[\ve{\radFlowNew},\ve{\prodInputNew}\right]}\in \mathbb{R}^{\card{\Eop}} $ and the new set of state constraint $\tilde{\inEqualCon}(\pentopVar,\designVarNew,\stateVar)$, constituting the adapted optimization problem:

\begin{equation} \label{eq:optProblemNew}
	\begin{aligned} \min_{\pentopVar,\designVarNew,\stateVar}& \qquad
		\costFull \left(\pentopVar, \designVarNew,\stateVar\right) \\
		s.t.& \qquad \tilde{\equalCon}(\pentopVar,\designVarNew,\stateVar) = 0, \\
		& \qquad \tilde{\inEqualCon}(\pentopVar,\designVarNew,\stateVar) \leq 0,  \\
		& \qquad \topVarBoxConstraints{\pentopVar},  \\
		& \qquad \opVarBoxConstraints{\designVarNew}.
	\end{aligned}
\end{equation}

As is practice in PDE-constrained optimization, the optimization problem \ref{eq:optProblemNew} is not solved directly, because it would require the optimization of both design variables $\pentopVar$, $\designVarNew$ and state variable $\stateVar$. To avoid the costly exploration within the feasible region of the physical model, we enforce that $\stateVar\left(\pentopVar,\designVarNew\right)$ is a solution to the system of non-linear model equations $\tilde{\equalCon}(\pentopVar,\designVarNew,\stateVar\left(\pentopVar, \designVarNew\right)) = 0$. This leads to a reduced optimization problem: 
\begin{equation} \label{eq:optReduced}
	\begin{aligned} \min_{\pentopVar,\designVarNew}& \qquad
		\hat{\costFull} \left(\pentopVar, \designVarNew\right)=\costFull \left(\topVar, \designVarNew,\stateVar\left(\pentopVar, \designVarNew\right)\right) \\
		s.t.& \qquad \hat{\inEqualCon}(\pentopVar,\designVarNew)=\tilde{\inEqualCon}(\pentopVar,\designVarNew,\stateVar\left(\pentopVar, \designVarNew\right)) \leq 0,  \\
		& \qquad \topVarBoxConstraints{\pentopVar},  \\
		& \qquad \opVarBoxConstraints{\designVarNew}.
	\end{aligned}
\end{equation}

\subsection{Optimization using augmented Lagrangian approach}\label{sec:AugLag}
To solve the reduced optimization problem in equation \ref{eq:optReduced}, an Augmented Lagrangian approach is proposed. In the previous paper published by the authors \cite{Blommaert2020b}, a Sequential Quadratic Programming (SQP) algorithm was used to include state constraints. This method has the drawback that it requires one gradient evaluation per constraint and the convergence of SQP solvers can be sensitive to infeasible starting points.

To avoid this, the algorithm was adapted to an Augmented Lagrangian approach. For this, equation \ref{eq:optReduced} is first reformulated as an equality-constrained problem by introducing a slack variable $\ve{\slack}$: $\hat{\inEqualCon}(\pentopVar, \designVarNew) + \ve{\slack}:= \hat{\ve{\equalConState}}\left(\pentopVar, \designVarNew,\ve{\slack}\right)=0$ and $
\ve{\slack} \geq 0$. The optimization problem is then solved by solving a series of subproblems with increasing constraint penalization $\LagPen$:

\begin{equation} \label{eq:optSlack}
	\begin{aligned} \min_{\pentopVar,\designVarNew,\ve{\slack}}& \qquad
		\ALagrangian\left(\pentopVar,\designVarNew,\ve{\slack},\ve{\LagMultis};\LagPen\right) \\
		s.t.& \qquad\ve{\slack} \geq 0\, ,\\
		& \qquad \topVarBoxConstraints{\pentopVar},  \\
		& \qquad \opVarBoxConstraints{\designVarNew}.
	\end{aligned}
\end{equation}
where the equality constraints are incorporated into an augmented Lagrangian:
\begin{align}
	\ALagrangian\left(\pentopVar,\designVarNew,\ve{\slack},\ve{\LagMultis};\LagPen\right) &= \hat{\costFull}\left(\pentopVar, \designVarNew\right) - \sum_{k =1 }^{m} \LagMultis_{k}\hat{\equalConState}_{k}\left(\pentopVar, \designVarNew,\ve{\slack}\right) \\
	&+ \frac{\LagPen}{2}\sum_{k = 1}^{m}\hat{\equalConState}^2_{k}\left(\pentopVar, \designVarNew,\ve{\slack}\right) \, .
\end{align}
The resulting optimization sub-problems then reduce to bound-constrained subproblems. Similar to Blommaert et al. \cite{Blommaert2020b}, these subproblems are solved using an SQP approach in which the gradient $\nabla\ALagrangian$ is computed using the discrete adjoint method. Hessian information is retrieved using a BFGS algorithm. It should be noted that in each iteration only a single adjoint calculation of the augmented Lagrangian is needed, in contrast to directly applying the SQP approach, for which an additional adjoint gradient is needed for each state constraint. Once this subproblem has been approximately solved, the multipliers $\ve{\LagMultis}$ and the penalty parameter $\LagPen$ are updated and the process is repeated \cite[p.520]{Nocedal2006}.

\subsection{Continuation and smoothing}
Applying a diameter penalization of $\pen>0$ introduces a multitude of additional local optima into the optimization problem. To avoid getting stuck prematurely in these local optima, a numerical continuation strategy is employed, gradually forcing the optimization to discrete values. In this approach, a series of optimizations is run, each using the optimum of the previous run as an initial guess. In this way, the penalization parameter is slowly increased in every continuation step following $\pen = \{0,2,4\}$.

The penalization also introduces non-differentiabilities at the desired available diameters (see figure \ref{fig:penInv}, \ref{fig:penHydr} and \ref{fig:penTherm}) into the optimization problem, which pose an additional challenge for the use of gradient-based optimization algorithms. To alleviate this problem, the non-smoothness originally introduced by the $\min$ function in equation \ref{eq:penalization}, is eliminated using a smooth approximation.\footnote{The $\min$ function can be reformulated as $\min\left(a,b\right)=-\relu\left(-a+b\right)+b$ using the rectifier function $\relu\left(x\right) = \max\left(x,0\right)$. This rectifier function is then smoothly approximated using the Gaussian Error Linear Unit function $\relu\left(x\right) \approx x \Theta\left(x\right)$ by Hendrycks and Gimpel \cite{Hendrycks2016}.}

\section{Demonstration on an academic optimal heat network design problem}
In this section, the topology optimization algorithm is tested on an academic heat network problem. First,  the correct convergence of the augmented Lagrangian treatment of the state constraints towards feasibility is verified in section \ref{sec:resState}. Then, the importance of a detailed economic problem formulation is discussed in section \ref{sec:resEcon}. Finally, the correct functioning of the novel multi-material topology optimization algorithm for heat network optimization is analysed in section \ref{sec:resTopOpt}. 

\subsection{Case set-up}\label{sec:setuo}
\begin{figure}
	\includegraphics[width=1\columnwidth]{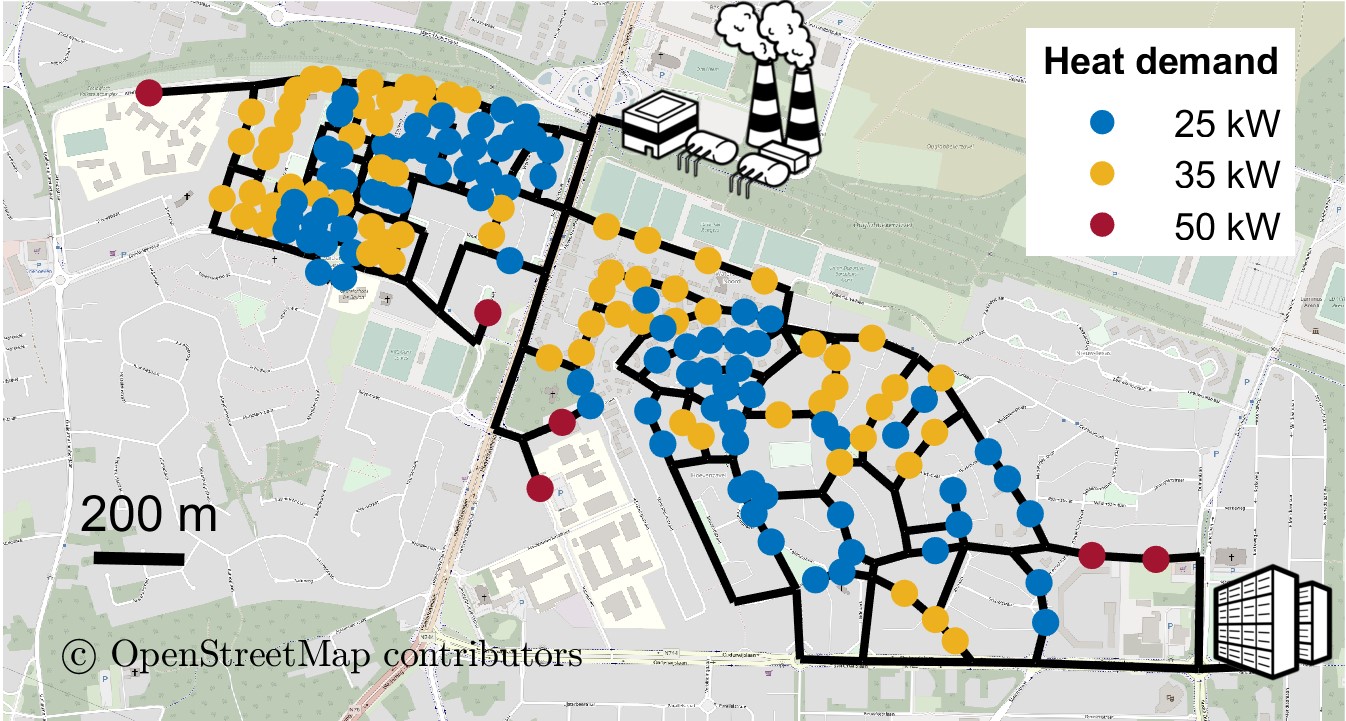}	
	\caption{Initial superstructure for the topology optimization problem for a district heating network test case. It connects two heat sources (north \& south-east) with 160 consumers represented by coloured circles.}
	\label{fig:setup}
\end{figure}

To test the topology optimization algorithm, an academic test case is set up. Here, a DHN is planned for a neighbourhood in Genk, Belgium. In this neighbourhood, 160 potential heat consumers  of varying heat demands ($\Qdemandi{\gi\gj}\in\{\SI{25}{\kilo\watt},\SI{35}{\kilo\watt},\SI{55}{\kilo\watt}\}$) are to be connected to two heat suppliers. A high temperature heat source at $\dTinf = \SI{70}{\degreeCelsius}$ (e.g. a gas power plant) and a low temperature source at $\dTinf = \SI{55}{\degreeCelsius}$ (e.g. a waste heat source like a data centre). Designing the optimal heat network that connects demand and supply, is posed as a topology optimization problem, using the neighbourhoods street grid as a superstructure. The set-up is illustrated in figure \ref{fig:setup}.

The properties within the optimization problem \ref{eq:optSlack} for this test case are summarized in table \ref{tab:properties}. 

\begin{table}[h!]
	\centering
	\caption{Properties of the topology optimization problem used in this test case. }	
	\label{tab:properties}
	\begin{tabular}{c|c|c}		
		Property & Value & Unit\\
		\hline
		$\cHeatCAPEXi{\gi\gj} \in$ &  $\{1000,0\}$ & $\unit{\sieuro\per\kilo\watt}$ \cite{Mertz2016}\\
		$\cHeatOPEXi{\gi\gj} \in$ &  $\{0.01,0.01\}$ &$\unit{\sieuro\per{\kilo\watt\hour}}$ \cite{Mertz2016}\\
		$\cPumpCAPEX$ & 
		$100$ & $\unit{\sieuro\per{\kilo\watt}}$\\
		$\cPumpOPEX$ &  $0.11$ & $\unit{\sieuro\per{\kilo\watt\hour}}$\\
		$\prodInputi{\gi\gj} \in$ &  $\{70,55\}$& $\unit{\degreeCelsius}$\\
		$\density$ &  $983$ & $\unit{\kilogram\per\meter^3}$\\
		$\viscosity$ &  $4.67\times10^{-4}$ & $\unit{\pascal\second}$\\
		$\spHeatCap$ &  $4185$ & $\unit{\joule\per{\kilogram\kelvin}}$\\
		$\TOutside$ &  $-8$ & $\unit{\degreeCelsius}$\\
		$\condGround$ &  $1$ & $\unit{\watt\per\meter\kelvin}$\\
		$\condInsul$ &  $0.0225$ & $\unit{\watt\per\meter\kelvin}$\\
		$\depthPipe$ &  $1$ & $\unit{\meter}$\\
		$\ratioInsul$ &  $1.87$ & $1$\\
		$\pumpEff$ & $0.81$ & $1$\\
		$\cRev$ & $0.08$ & $\unit{\sieuro\per{\kilo\watt\hour}}$ \\ 
		$\Qdemandi{\gi\gj} \in $ &$\{25,35,55\}$&$\unit{\kilo\watt}$ \\
		$\gijEdge{\heaterCoef} \in $ &$\{450,650,1000\}$&$\unit{\watt\per\kelvin^\heaterExp}$ \\	
		$\gijEdge{\heaterExp} \in $ &$\{1.42,1.2,1.2\}$&$1$ \\	
		$\setDiscreteDiameters$ & $ \{1,3,7,11,15,20\}$ & $\unit{\centi\meter}$ \\
		$\costDiscreteDiameters$ & $400+\{180,195,234,\dots$& $\unit{\sieuro\per\meter}$ \\
		&$275,400,461\}$&\\
	\end{tabular}
\end{table}

\subsection{Results}\label{sec:results}
The topology optimization problem is now solved using the above mentioned algorithm. In this first part of the analysis continuous pipe diameters are tolerated, so the proposed penalization method of section \ref{sec:penalization} is not yet used ($\pen = 0$). The result with an active penalization are discussed in section \ref{sec:resTopOpt}. 

\begin{figure}[h]
	\includegraphics[width=1\columnwidth]{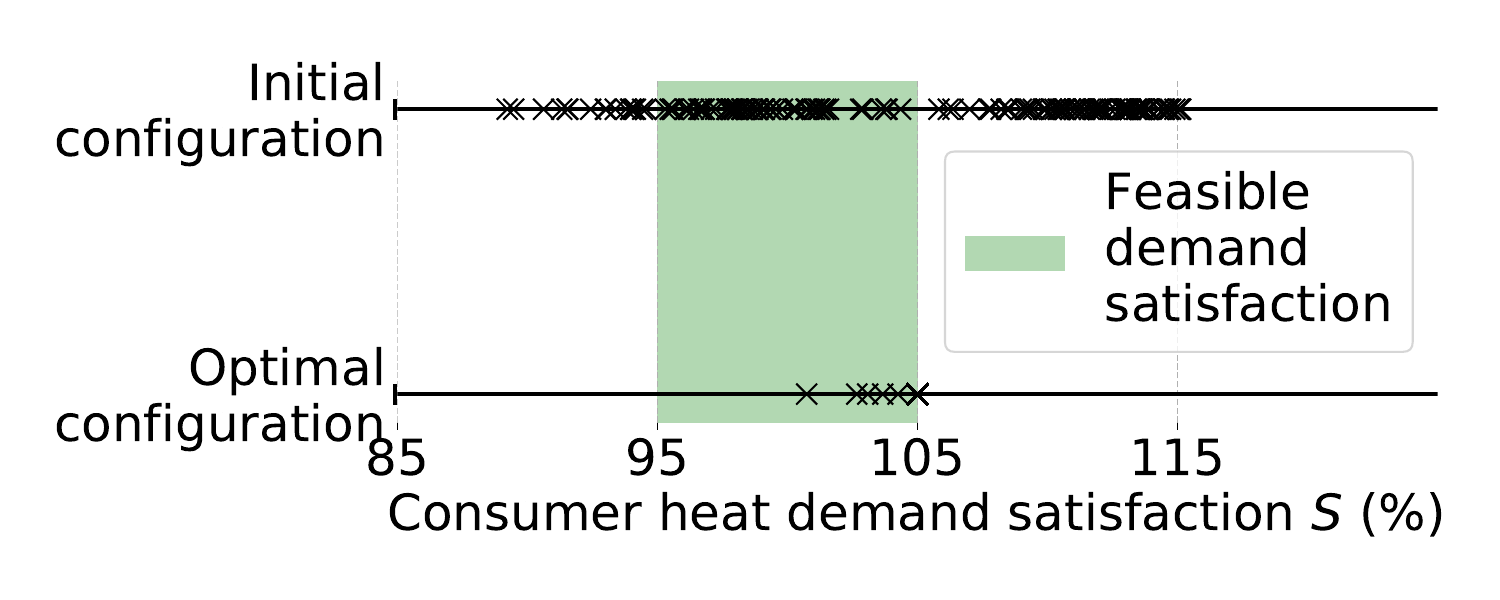}
	\caption{Heat demand satisfaction $\DemandSatisfaction$ of all consumers before (top) and after the optimization (bottom). The feasible region defined by constraint \ref{eq:heatSatisfaction} is shown in green.}
	\label{fig:demandSatis}
\end{figure}

\subsubsection{Correct state constraint treatment}\label{sec:resState}
First, the correct treatment of state constraints with the Augmented Lagrangian approach is verified. To this end, the heat demand satisfaction  $\DemandSatisfaction = \left(\frac{\gijEdge{Q}-\Qdemandi{\gi\gj}}{\Qdemandi{\gi\gj}}\right)$, is plotted for all consumers both at the beginning and end of the optimization in figure \ref{fig:demandSatis}. It can be observed that the heat demand of all consumers is indeed met. The Augmented Lagrangian approach is therefore indeed able to enforce state constraints without the need for an additional warm-start (As compared to Blommaert et al. \cite{Blommaert2020b}).

\subsubsection{On the importance of an economic cost function}\label{sec:resEcon}
\begin{figure}[h]
	\includegraphics[width=1\columnwidth]{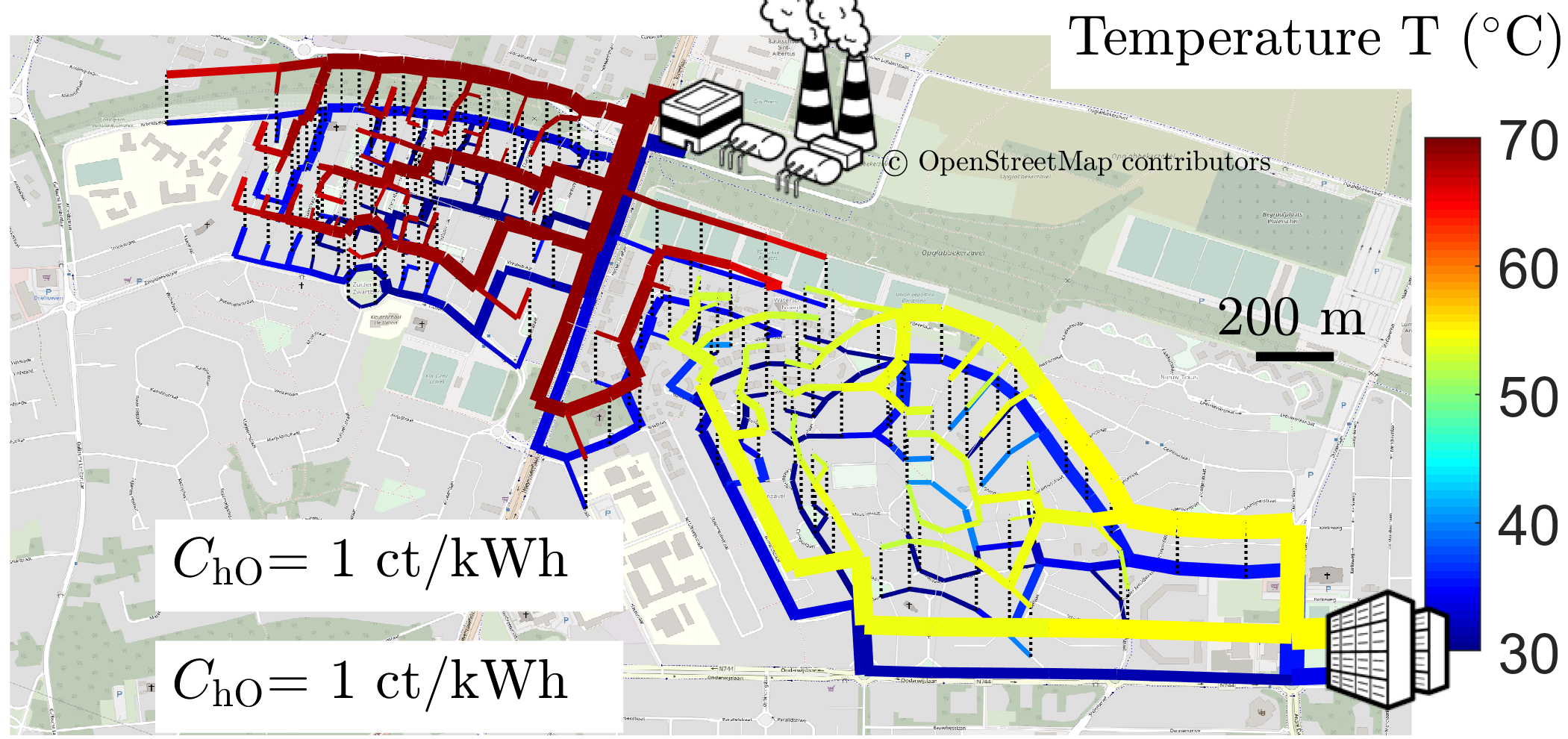}	
	\caption{Optimal heat network topology. Here, both feed (top) and return pipes (bottom) are shown. Consumers are represented by dotted lines connecting feed and return network. Heat production facilities are represented by icons. The optimal pipe diameter is shown with the line-thickness, while the line-colour represents the water temperature within the pipe.}
	\label{fig:economicResult1}
\end{figure}

Now the resulting optimal network topology is plotted in figure \ref{fig:economicResult1}.

It can be observed that the optimal heat network topology for this case contains two individual networks. One provided by the northern producer at a high temperature ($\approx\SI{70}{\degreeCelsius}$), the other by the south-eastern producer at a lower temperature ($\approx\SI{55}{\degreeCelsius}$). In contrast to typical applications of PDE-constrained topology optimization,  where the cost function is often chosen as a physical quantity that is to be minimized, such as the structural compliance in structural optimization problems (see Bendsøe and Sigmund \cite{Bendsoe2003}) or the mean temperature in heat transfer optimization problems  (e.g. Yu et. al. \cite{Yu2020}), the net present value is directly maximized here. The net present value of the optimized design amounts to $\npv = \SI{19.157}{\mega\sieuro}$.

\begin{figure}[h]
	\includegraphics[width=1\columnwidth]{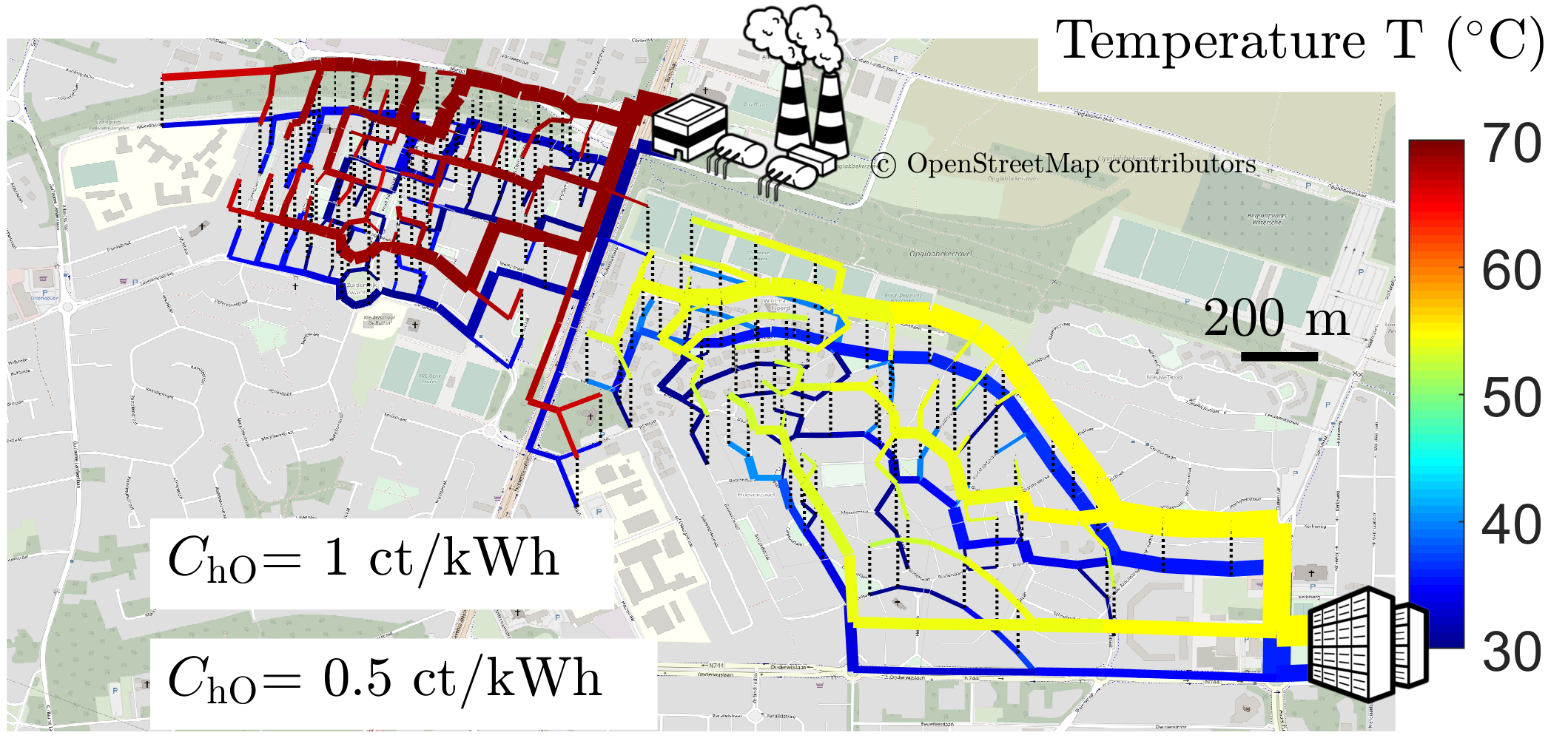}	
	\caption{Shift of network topology when lowering the heat acquisition price of the waste heat source to $\cHeatOPEX = \SI{0.5}{\ct\per{\kilo\watt\hour}}$. More houses are now connected to the low temperature source.}
	\label{fig:economicResult2}
\end{figure}

For DHNs, it is beneficial to take the economic perspective. Indeed, topology optimization here aids in the investment decision of a complex energy system. Moreover, it has the advantage of allowing to use a topology optimization strategy to study the influence of economic parameters (e.g. the producer heat price $\cHeatOPEX$) on the final cost of the network. To highlight this, a design study is done on the heat OPEX parameter $\cHeatOPEX$ of the waste heat source in the south-east. To evaluate the influence of a different heat pricing scenario, the heat acquisition price of this source is decreased to $\cHeatOPEX = \SI{5}{\ct\per{\kilo\watt\hour}}$. When connecting a waste heat source, this can be acceptable as this heat is otherwise unused. The optimal network topology for this scenario can be seen in figure \ref{fig:economicResult2}.

It is apparent, that the topology shifted and more houses are now connected to the low temperature waste heat source as this heat can be acquired at a lower cost. This is also visible in the increase in $\npv$ of the network to $\SI{24.267}{\mega\sieuro}$. The limit to the size of the low-temperature waste heat network is given by the temperature-dependent efficiency of the consumer heating systems. The fact that the optimal configuration results from a balance between heat acquisition cost and heating system efficiency illustrates well the importance of combining an extensive economical analysis with a detailed physics model of the heat network. 

\subsubsection{Near-discrete pipe design}\label{sec:resTopOpt}
\begin{figure}[h]
	\includegraphics[width=1\columnwidth]{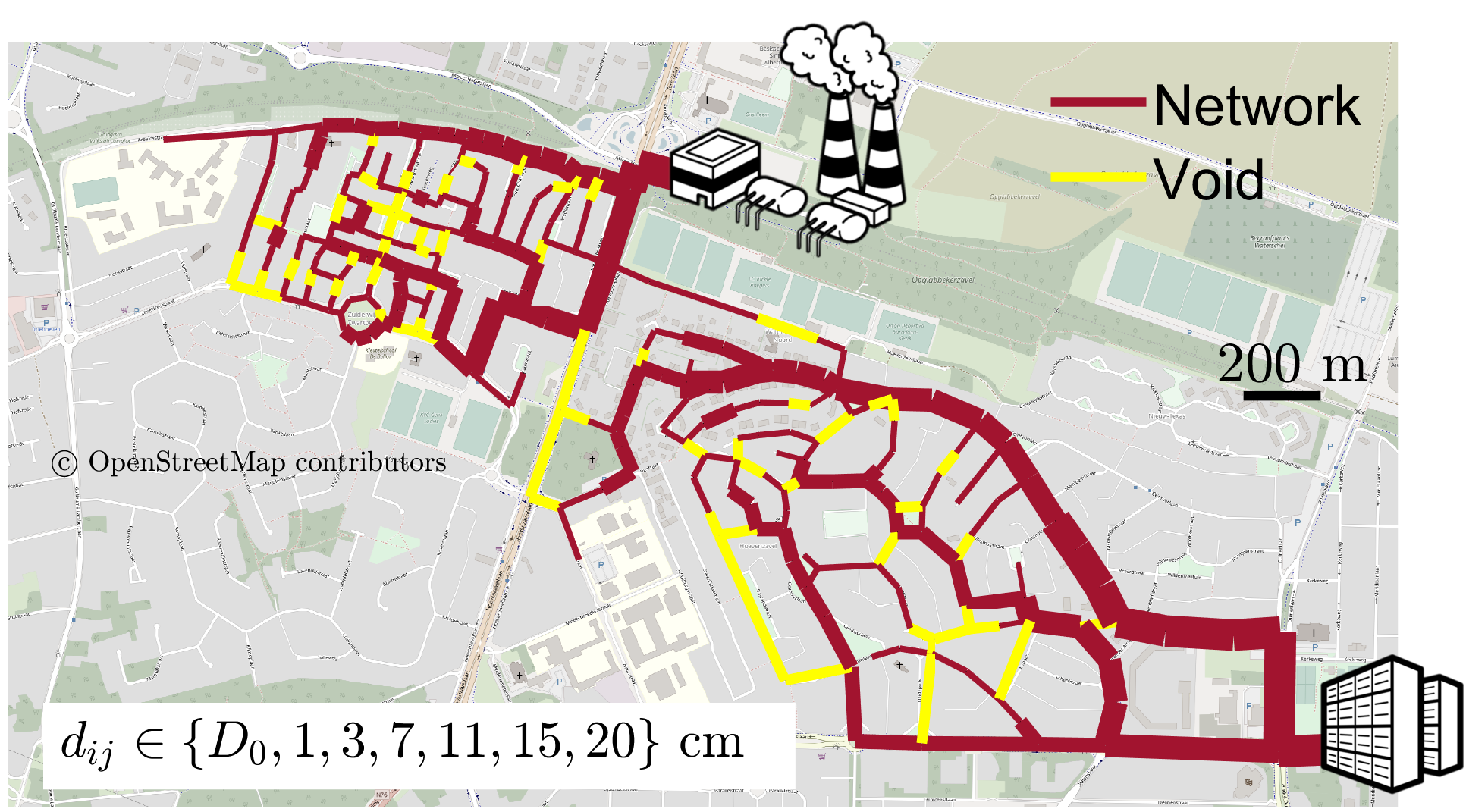}
	\caption{Discrete DHN topology and pipe design after using the pipe penalization strategy. The pipes are chosen from $\diameter_{\gi\gj} \in \{\diameterDiscrete_{0},\SI{1}{\centi\meter},\SI{3}{\centi \meter}, \SI{7}{\centi \meter}, \SI{11}{\centi \meter}, \SI{15}{\centi \meter}, \SI{20}{\centi \meter}\}$, void (no pipe) is represented by a yellow line.}
	\label{fig:discTopology}
\end{figure}

In order to build an optimal heat network, near-discrete pipe design has to be achieved. For this, the pipe penalization strategy from section \ref{sec:penalization} is applied to the test case of section \ref{sec:setuo} with a set of available pipe diameters of $ \{\diameterDiscrete_{0},\SI{1}{},\SI{3}{}, \SI{7}{}, \SI{11}{}, \SI{15}{}, \SI{20}{}\}\unit{\centi \meter}$. The resulting optimal topology and near-discrete pipe diameter is visualized in figure \ref{fig:discTopology}. It can be seen that in contrast to the previous continuous optimizations (figure \ref{fig:economicResult1} \& \ref{fig:economicResult2}), distinct discrete jumps between the available discrete diameters are visible.

\begin{figure}[h]
	\includegraphics[width=1\columnwidth]{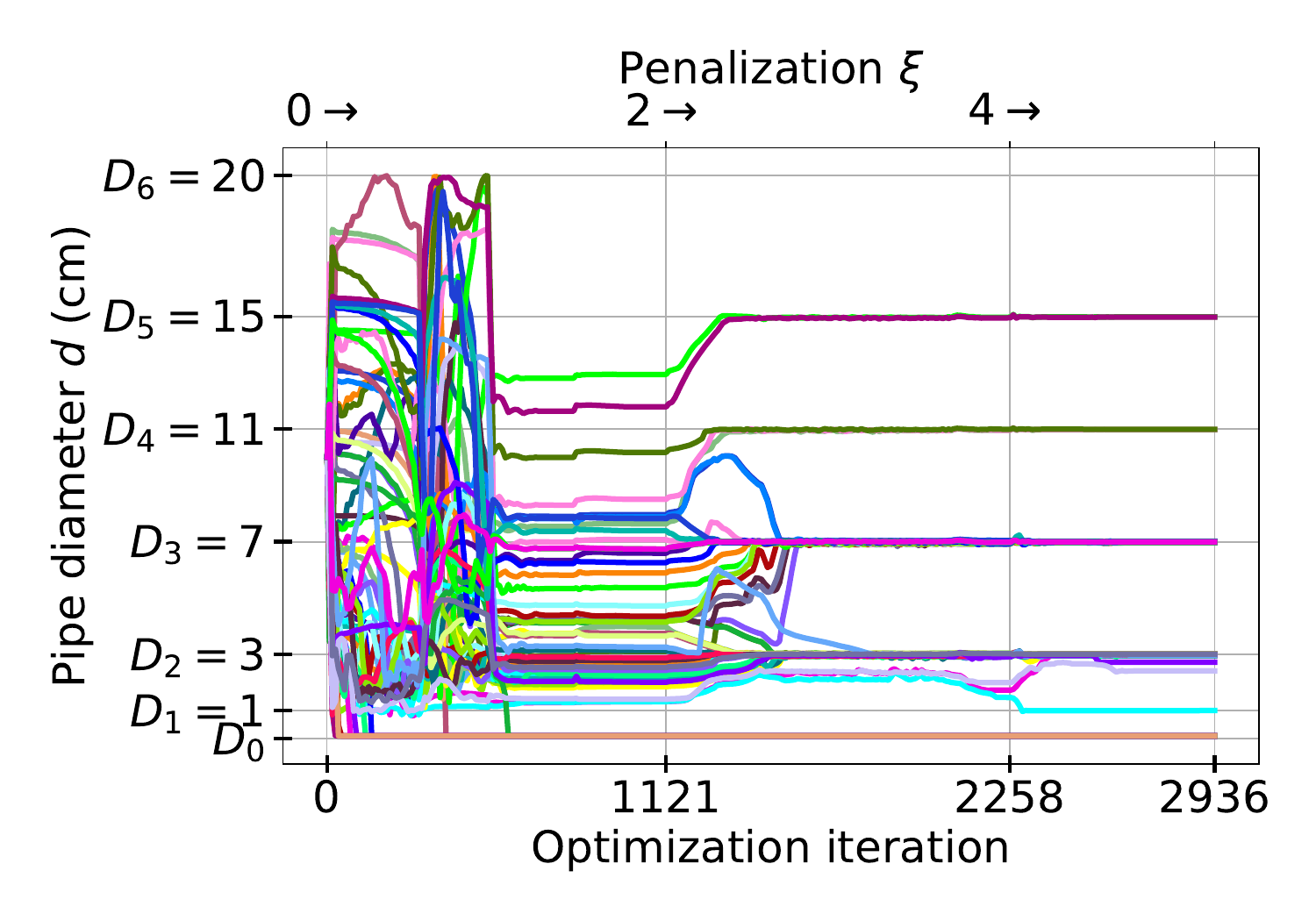}
	\caption{Evolution of a random sample of pipe diameters $\gijEdge{\diameter},\, \defSetPipes$ in the heat network with the optimization iterations. The increasing penalization $\pen \in \{0,2,4\}$ is plotted in the upper abscissa. Pipe diameters converge towards discrete values for increasing penalization values $\pen$. The full plot is shown in Appendix \ref{sec:fullPlot}.}
	\label{fig:designEvol}
\end{figure}

To show that near-discrete design was indeed achieved, the evolution of pipe diameters in the network over the optimization iterations is plotted in figure \ref{fig:designEvol}. Additionally the continuation steps on the penalization parameters are plotted on the upper abscissa. It can be seen that starting from wide distribution of diameters for $\pen=0$, the diameters tend towards discrete values with increasing penalization ($\pen=2$ \& $\pen=4$). The design evolution plot also unveils that a few non-discrete (``grey'') variables remain after the final penalization step  (between the discrete pipe sizes of 1 and 3cm). These grey design variables are a common phenomenon in topology optimization and further steps could be taken to eliminate them (e.g. bu further increasing the penalization or by using the Heavyside Projection method by Guest \cite{Guest2009}).

\begin{figure}[h]
	\begin{center}
		\includegraphics{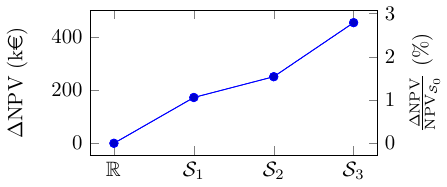}
%
%
%
	\end{center}
	\begin{center}
		\begin{tabular}{c|ccc}		
			
			$\gijEdge{\diameter}\in$&$\setDiameters_1 $  & $\setDiameters_2$ & $\setDiameters_3$   \\			
			\hline
			&\multicolumn{3}{c}{$\npv(\unit{\kilo\sieuro})$}\\\hline
			Penalization& $\SI{15612}{}$ & $\SI{15937}{}$ & $\SI{14856}{}$ \\
			
			Rounding up& $\SI{15440}{}$ & $\SI{15687}{}$ & $\SI{14404}{}$\\
			\hline
			Difference  &\multirow{2}{*}{$\SI{172}{}$}  &\multirow{2}{*}{$\SI{250}{}$}  & \multirow{2}{*}{$\SI{453}{}$}\\
			($\Delta \npv$) & \\
		\end{tabular}
	\end{center}
	\caption{$\npv$ of the optimal heat network topology for different sets of available pipe diameters $\setDiameters_k$ ($\setDiameters_1 = \{1,3,7,11,15,20\}\,\unit{\centi\meter},\,\setDiameters_2=\{3,7,15,20\}\,\unit{\centi\meter},\, \setDiameters_3=\{3,11,20\}\,\unit{\centi\meter}$). Here, the proposed penalization method is compared to a simple rounding up post-processing. The penalization method increasingly outperforms "rounding" with an increasing scarcity $\card{\setDiameters_k}$ of the available pipe diameters.}
	\label{fig:discretDiameters}
\end{figure}

The $\npv$ of the discrete network is, as expected, with $\npv=\SI{14.612}{\mega\sieuro}$ lower than the $\npv = \SI{16.227}{\mega\sieuro}$ of an optimal network with continuous pipe design. To evaluate if the novel pipe penalization strategy leads to better discrete designs then a simple post-processing step (e.g. rounding up of the continuous design), a comparative study is conducted. Here, the $\npv$ of the optimal discrete network design is compared to the $\npv$ of a rounded design, starting from the optimal continuous diameters. The results of this study, for multiple discrete sets of available pipe diameters ($\setDiameters_1 = \{1,3,7,11,15,20\}\,\unit{\centi\meter},\,\setDiameters_2=\{3,7,15,20\}\,\unit{\centi\meter},\, \setDiameters_3=\{3,11,20\}\,\unit{\centi\meter}$), can be seen in figure \ref{fig:discretDiameters}.\footnote{Note the counter-intuitive increase in the $\npv$ from $\setDiameters_1$ to $\setDiameters_2$. This can be explained by the non-convex nature of the optimization problem, because of which we can only guarantee convergence to a local optimum. In this case, it is clear that a better local optimum can be found for $\setDiameters_1$, namely the solution of $\setDiameters_2$ and $\setDiameters_3$.} It is noted that for a clean cost comparison, the grey designs of the topology optimization are also rounded to the next pipe diameter.

First, an improvement of $\SI{172}{\kilo\sieuro}$ over simple rounding-up can be achieved for pipe catalogue $\setDiameters_1$.  The study shows that an improvement can be achieved for all three tested pipe catalogues $\setDiameters_k $ and that the magnitude of improvement increases with the scarcity of that catalogue $\card{\setDiameters_k}$. If the optimization is constrained to only three available pipe diameters $\setDiameters_1$, an improvement of $\SI{453}{\kilo\sieuro}$ was achieved over simple rounding, amounting to a relative improvement of $2.8 \%$ in reference to the $\npv_{\setDiameters_0}$ of the optimization case allowing for continuous pipe diameters. This shows that the newly introduced pipe penalization strategy has added value for topology optimization of heat networks and could significantly reduce the investment cost of future DHN development projects.
Such reduction is of major importance. Despite having great long term economic and ecological potential, DHN project feasibility is often hampered by the high initial cost compared to competed technologies. Piping infrastructure, and more specifically material cost, can account for up to 60\% of the total cost in the early stages, clearly illustrating the potential impact and gains of an optimized approach.

\section{Conclusion}
In this paper, we proposed a pipe diameter penalization strategy for the topology and pipe diameter optimization of heating networks. The penalization method efficiently produces optimal network topologies and near-discrete pipe designs for the economic optimization of a medium-sized District Heating Network project, without resorting to combinatorial optimization. The resulting discrete network designs has a consistently higher net present value then discrete networks designed using a simple post-processing step. This gain in net present value from using the topology optimization approach increases with the scarcity in available pipe diameter choices. A maximum improvement in net present value of $2.8 \%$ was achieved for an optimal design with only three available pipe diameters.  

In addition, the optimization problem was reformulated as an economical problem taking an investors perspective. Since the upfront investment costs are a primary factor for assessing district heating network design, a methodology is presented that allow to directly optimize the full net present value assessment of the heating network. It was shown for a medium sized District Heating Network project how economic parameters influence the optimal network topology. A decrease in the heat acquisition price of a waste heat source increased the amount of consumers connected to that source, ultimately increasing the share of waste heat in the optimized network. This case study highlights the potential of economic topology and design optimization based on a detailed physics model especially in the early design phases of District Heating Networks. Next, an Augmented Lagrangian approach for the treatment of state constraints, like satisfying the heat demand for all consumers, was introduced. The approach manages to satisfy the state constraints without the need for warm-starting the optimization.

This novel application of topology optimization methods to optimal District Heating Network design proves to be a valuable alternative to common combinatorial approaches in the field. The method is shown to produce near-discrete optimal network topologies and pipe designs, while maintaining physical accuracy with non-linear network models. Further research should aim at a detailed comparison of the performance of the topology optimization approach to that of combinatorial approaches. 

Finally, a fast and accurate optimization framework as presented in this work would allow to accurately perform scenario analysis of District Heating Networks on large scale and provide much needed support towards (decentralized) energy network planning. For a next step, the economic impact of large-scale non-linear optimization on real world District Heating Network project will therefore be studied. Further research should be conducted on reducing remaining grey pipe design, and on further improving the detail of the heating network model. Discrete pipe diameter optimization could be extended to include insulation, pipe material, and market mechanisms; e.g. mass-produced pipes with limited diameter choices versus tailored solutions.

\appendix
	
	\section{A partially-reduced space reformulation to facilitate convergence}\label{secA1}
	Introducing a high penalization $\pen$  in initial optimization stages, causes an ill-conditioning of the heat transport problem. The steep increase in the hydraulic resistance of pipes through the penalization (see figure \ref{fig:penHydr}) leads to diminishing flow rates throughout the network during the initialization. To avoid this ill-conditioning hindering convergence of the optimization problem, a reformulation of the initial optimization problem (equation \ref{eq:optContinuous}) as a partially-reduced space formulation is proposed.
	
	In contrast to the state-of-the-art in district heating topology optimization, our approach is based on a consistent set of model equations and boundary conditions in a reduced-space approach. However, to alleviate the ill-posedness of the optimization problem caused by the penalization of intermediate diameters, we propose a reformulation here in a partially-reduced space so that the model equations will only be satisfied at convergence of the KKT conditions for optimality. 
	
	First, we substitute the momentum equations (equation \ref{eq:radiatorMomentum}) of the consumer edges in the box constraints $\opVarBoxConstraints{\designVar}\in\mathbb{R}^{\card{\Econ}} $, which yields for $\gijEdge{\valveRhydr}\gijEdge{\flow} \geq 0 $:
	\begin{align}
		&0 \leq \gijEdge{\radValve}= \frac{\gijEdge{\valveRhydr}\gijEdge{\flow}}{\giNode{\pressure} - \gjNode{\pressure}} \leq 1 \quad \defSetRad, \\
		\Leftrightarrow \quad&\gijEdge{\valveRhydr}\gijEdge{\flow}-\left(\giNode{\pressure} - \gjNode{\pressure}\right):=\inEqualCon_{\mathrm{m}}\left(\stateVar\right)  \leq 0, 
	\end{align}
	To again close the system of model equations a boundary condition for the consumer arcs is defined as
	\begin{equation}
		\gijEdge{\flow}-\gijEdge{\radFlowNew}\flow_{\mathrm{max},\gi\gj}=0\, , \quad\defSetCon,
	\end{equation}
	and the producer inflow is replaced by a pressure driven boundary condition:
	\begin{equation}
		\giNode{\pressure}-\gjNode{\pressure} = \giNode{\prodInputNew}\quad \forall i \in \NproF,\forall j \in \NproR\,.
	\end{equation}

	This leads to a new system of model equations $\tilde{\equalCon}\left(\pentopVar,\designVarNew,\stateVar\right) = 0$ with the new design variable vector $\designVarNew = \tp{\left[\ve{\radFlowNew},\ve{\prodInputNew}\right]}\in \mathbb{R}^{\card{\Eop}} $. 
	This significantly simplifies the model equations, eliminating the ill-conditioning. By substituting the consumer momentum equation into the bound constraints for the valve settings, these constraints $\inEqualCon_{\mathrm{m}}\left(\stateVar\right)$ now have to be treated as state constraints and are therefore combined with the vector of generic state constraints $\inEqualCon(\pentopVar,\designVarNew,\stateVar)$ in $\tilde{\inEqualCon}(\pentopVar,\designVarNew,\stateVar)=\tp{\left[\inEqualCon(\pentopVar,\designVarNew,\stateVar),\inEqualCon_{\mathrm{m}}\left(\stateVar\right)\right]}$, constituting the adapted optimization problem:
	\begin{equation} 
		\begin{aligned} \min_{\pentopVar,\designVarNew,\stateVar}& \qquad
			\costFull \left(\pentopVar, \designVarNew,\stateVar\right) \\
			s.t.& \qquad \tilde{\equalCon}(\pentopVar,\designVarNew,\stateVar) = 0, \\
			& \qquad \tilde{\inEqualCon}(\pentopVar,\designVarNew,\stateVar) \leq 0,  \\
			& \qquad \topVarBoxConstraints{\pentopVar},  \\
			& \qquad \opVarBoxConstraints{\designVarNew}.
		\end{aligned}
	\end{equation}

	\section{Full plot of the design evolution of every pipe diameter} \label{sec:fullPlot}
	
	\begin{figure}[h!]
		\includegraphics[width=1\columnwidth]{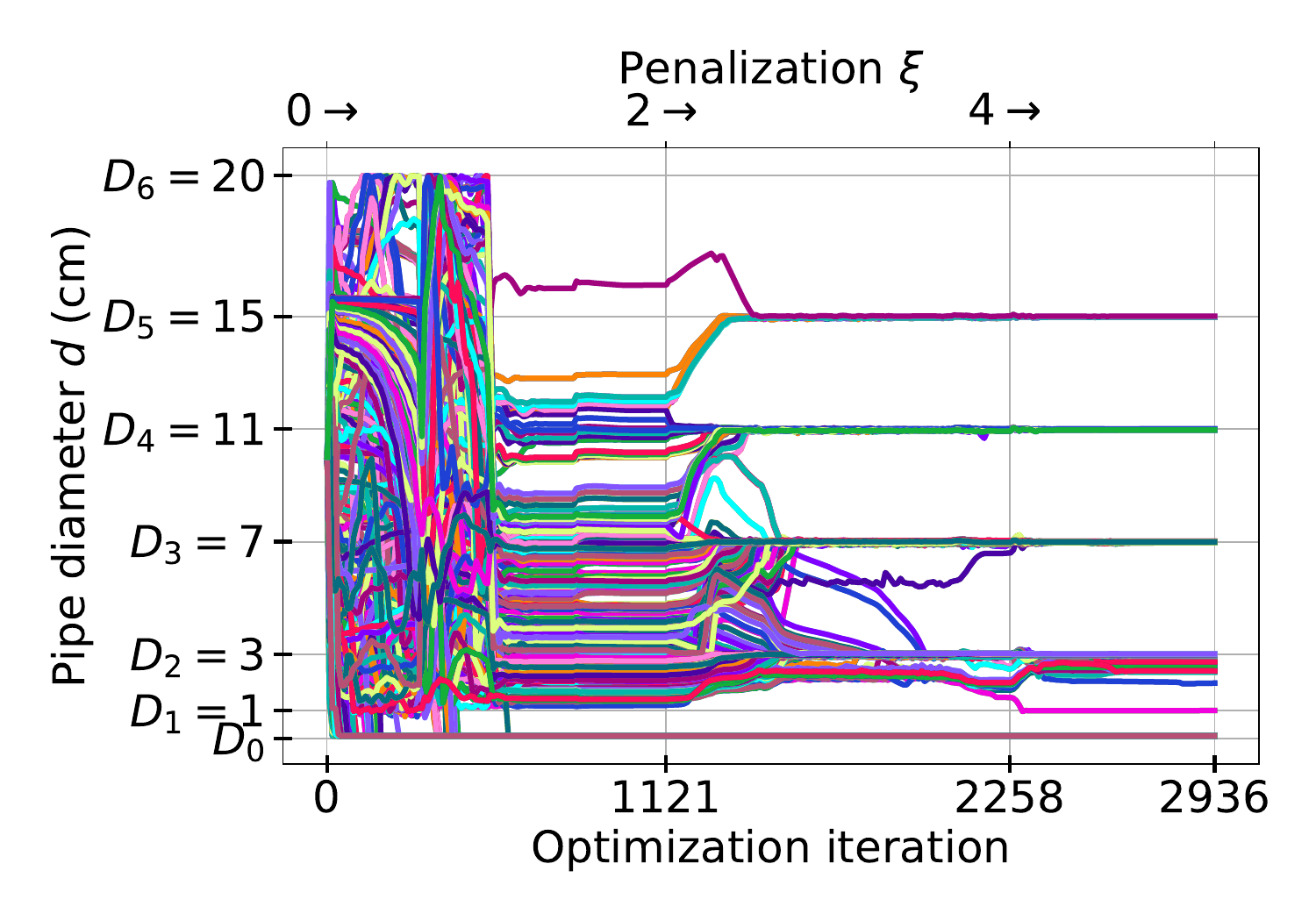}
		\caption{Evolution of every pipe diameter $\gijEdge{\diameter},\, \defSetPipes$in the heat network with the optimization iterations. The increasing penalization $\pen \in \{0,2,4\}$ is plotted in the upper abscissa. Pipe diameters converge towards discrete values for increasing penalization values $\pen$.}
		\label{fig:designEvolFull}
	\end{figure}

\section*{Data Availability}
A data-set including the structure, input parameters and optimization results of the heating network used in the test case of this paper is available at the following link: \url{https://doi.org/10.48804/56GXSC}. The optimization results can be replicated using the methodology and formulations described in this paper.

\section*{Aknowledgements}
Yannick Wack is funded by the Flemish institute for technological research (VITO). Robbe Salenbien is supported via the energy transition funds project ‘EPOC 2030-2050’ organized by the FPS economy, S.M.E.s, Self-employed and Energy.

\section*{CRediT authorship contribution statement}
\textbf{Yannick Wack}: Conceptualization, Methodology, Software, Visualization, Writing – original draft. \textbf{Tine Baelmans}: Conceptualization, Funding acquisition, Writing – review \& editing. \textbf{Robbe Salenbien}: Conceptualization, Funding acquisition, Writing – review \& editing. \textbf{ Maarten Blommaert}: Conceptualization, Methodology, Software, Supervision, Funding acquisition, Writing – review \& editing 

\section*{Compliance with Ethical Standards}
\subsection*{Conflict of Interest}
The authors declare that they have no conflict of interest.
\subsection*{Funding}
The authors did not receive support from any organization for the submitted work.
\subsection*{Ethical approval}
This article does not contain any studies with human participants or animals performed by any of the authors.

\bibliography{Manuscript}

\end{document}